\documentclass[twocolumn,showpacs,amsmath,amssymb,letterpaper]{revtex4}

\usepackage{graphicx}
\usepackage{dcolumn}
\usepackage{bm}
\usepackage{hyperref} 
\usepackage{multirow} 

\begin{document}
\title{Particle Production in  $\sqrt{s_{\rm NN}}  = 2.76$ TeV  Heavy Ion Collisions}%
\author{Johann Rafelski$^1$}
\author{Jean Letessier$^{1,2}$}
\affiliation{%
$^1$Department of Physics, University of Arizona, Tucson, Arizona 85721, USA
}%
\affiliation{%
$^2$LPTHE,Universit{\' e} Paris 6 et 7, Paris 75005, France}%

\date{ December 7, 2010} 

\begin{abstract}
We consider, within the statistical hadronization model,  the near central 
rapidity $y\simeq 0$  integrated hadron yields  expected at LHC
 $\sqrt{s_{\rm NN}}  = 2.76$ TeV  ion reactions, for which the total charged 
hadron  rapidity most central head-on collision yield is  $dh/dy|_{y=0}\simeq 1800$\,.
For the chemical equilibrium SHM, we discuss composition of $dh/dy$
as function of hadronization temperature. For  chemical non-equilibrium SHM,
we input the specific strangeness yield $s/S$, demand explosive disintegration and
study the  break up as a function of the  critical hadronization pressure   $P$. 
We develop  observables  distinguishing the hadronization 
models and conditions.
\end{abstract}


\maketitle

{\bf Introduction:}
According to the theory of strong interactions,
Quantum Chromo-Dynamics~(QCD),  quarks 
and gluons are confined inside hadrons.
At sufficiently high temperature  lattice 
computations demonstrate that the deconfined  Quark--Gluon 
Plasma (QGP) prevails~\cite{Borsanyi:2010cj}.
We seek  to prove this QCD paradigm
in high energy heavy ion collisions where
heavy nuclei   are crushed on each other, forming
a small drop of thermalized deconfined QGP matter.
This drop   hadronizes into a high multiplicity
of particles and  we seek to determine the physical
properties of QGP considering   this multiparticle final
state.

At LHC energy  $\sqrt{s_{\rm NN}}  = 2.76$ TeV,
for near head-on   5\% most central Pb--Pb
collisions, the pseudo-rapidity density of primary 
charged particles at mid-rapidity  is
 $dh/d\eta={1584}  \pm  {4}  \pm   {76} $ {\it syst}, 
an increase of about a factor   {2.2} to central Au--Au
RHIC collisions at  
$\sqrt{s_{\rm NN}}  = 0.2$~TeV~\cite{Aamodt:2010pb}.
The study of particle yields per unit of rapidity 
obtained after integration of transverse 
momentum particle spectra eliminates the need to model
the distortion of  spectra introduced by 
explosive dynamics (see, e.g.,~\cite{Bozek:2010wt}) 
of highly compressed matter  created in high energetic 
collisions.

Following the reference~\cite{Bozek:2010wt}, where deformation
of the spectra were studied in model cases, we  interpret
the measured $dh/d\eta$ to be equivalent to a central rapidity 
density $dh/dy\simeq1800$.
This   experimental result  enables us to offer prediction  
for a variety of different  particle multiplicities, 
dependent on adopted  hadronization
model and a first interpretation 
 in terms of particle source bulk properties.

{\bf SHM --- Statistical hadronization model:}
QGP hadronic particle production yields are generally 
considered within the statistical hadronization 
model (SHM)~\cite{Torrieri:2004zz,Letessier:2005qe}.
SHM has been successful in describing   hadron production in
heavy ion collisions for different colliding systems and 
energies. Some view the SHM as a qualitative model and as such 
one is tempted to seek   {\it simplicity}   in an effort to obtain 
an estimate of the yields for all hadrons with just 
a small number of parameters~\cite{Andronic:2005yp,Andronic:2008gu,Becattini:2010sk}. 

Improving experimental precision,  along with physics motivation
based in qualitative dynamics of the hadronization process
has stimulated refinements involving a greater parameter set allowing
to control the  dynamically established yield of different quark flavors,
generally referred to as chemical \underline{non}-equilibrium 
SHM~\cite{Letessier:2005qe}.  This is achieved by 
introducing statistical occupancy  parameters $\gamma_i>1,\ i=q,\,s,\,c$,
where $s$ is the strange and $c$ the charm quark flavor. It 
can be assumed that   up and down quark yields $q=u,\,d$ are equally 
equilibrated. We will not discuss the charm flavor here.

Moreover, we are  interested in  precise description 
of the bulk properties of the particle source, 
such as size, energy and entropy content of the 
QGP fireball. This requires  precise  capability 
to extrapolate observed hadron yields to unobserved 
kinematic domains and unobserved particle types. This
is the case for chemical \underline{non}-equilibrium 
approach as demonstrated by the smooth systematic
behavior of physical observables as a function of collision 
conditions such as reaction energy~\cite{Letessier:2005qe}
or collision centrality~\cite{Rafelski:2004dp}.

With increasing collision energy, the baryon content  at 
central  rapidity decreases rapidly. It is expected that 
there remains  a small excess of matter over antimatter 
at central rapidity at LHC~\cite{Andronic:2009gj}. 
In SHM, this is governed by chemical parameters 
$\lambda_q,\ \lambda_s$ (equivalent to $\mu_B,\ \mu_S$, 
the baryon and strangeness chemical potentials of matter). 
Determination of values of 
$\mu_B,\ \mu_S$  must await additional experimental data. 
We present, for orientation of how the baryon--antibaryon yields vary,
results testing different values of  the expected potentials
choosing a fixed input value  $\lambda_q =1.0055$,   
which results for in  $\mu_B\simeq 2.9\pm0.3$\,MeV, $\mu_S\simeq 0.7\pm0.3$\,MeV,
and  $\mu_B\simeq 2.0\pm0.2$\,MeV, $\mu_S\simeq 0.45\pm0.05$\,MeV in case of 
equilibrium and non-equilibrium models, respectively.  

Note that since the number of
strange and antistrange quarks in hadrons has to be equal, 
$\lambda_s$, and thus  $\mu_S$ is determined
in order to satisfy this constraint. Another constraint is also implemented, the 
total charge per baryon has to be $Q/B=0.4$, since  the stopping
of electrically charged matter (protons) within a rapidity interval 
is the same as that of all baryonic matter (protons and neutrons). This is
achieved by a suitable tiny up-down quark asymmetry, a feature implemented 
in the SHARE suit of programs we are using~\cite{Torrieri:2004zz}. 

In the usual procedure of statistical hadronization modeling,
the particle yields are used to find best statistical parameters.
The SHARE suit of programs was written in a more flexible way
to allow also mixed a fit, that is a fit where a few particle yields 
can be combined with the given `measured' statistical parameters
to obtain best fit of other statistical parameters. When we were 
developing SHARE, this feature was created since a parameter such as
temperature could be measured using spectral shape and thus should
be not fitted again in the yield description but be used as an 
experimental input. To be general, this feature was extended to
all statistical parameters of the SHARE program. 

This feature allows us to  perform a  fit of the mix of statistical 
parameters and particle yields. Our procedure
has been outlined before~\cite{Rafelski:2008an} and is, here, 
applied for the first time including LHC experimental data. This allows us 
to predict within the chemical equilibrium and non-equilibrium SHM models
the differing pattern of particle production such that $dh/dy$ is fixed.

{\bf Importance of understanding chemical (non)-equilibrium:}
When chemical \underline{non}-equilibrium is derived from the particle 
yields, generating in the fit to data $\gamma_i\ne 1$, 
this suggests  a dynamical picture of  an explosively
expanding and potentially  
equilibrated QGP,  decaying rapidly into free streaming hadrons.
Without a significant re-equilibration, the (nearly) equilibrated QGP cannot 
produce chemically equilibrated hadron yield. 
The high intrinsic QGP entropy content explains why equilibrated QGP turns into  
chemically overpopulated (over-saturated) HG phase space. --- The fast breakup of QGP 
means that the emerging hadrons 
do not have opportunity  to re-establish chemical equilibrium in the HG phase.

The differentiation of chemical equilibrium and non-equilibrium SHM models will 
be one of the challenges we address in our present discussion
facing the SHM model and interpretation of the hadron 
production. One could think that resolution of this matter
requires a good fit of SHM parameters 
to the data. However, with the large errors on particle yields and lack of 
sensitivity to $\gamma_q$, this is not easy. 

The light quark phase space occupancy parameter 
 $\gamma_q$ can be only measurable by determining overall baryon to meson yield, and 
this cannot be done without prior measurement of hadronization temperature $T$.
When particle yield data is not available to measure both $T$  and $\gamma_q$, one can
fit only  $\gamma_s/\gamma_q$ to the data, which is then reported as  
$\gamma_s$ accompanied by the tacit assumption $\gamma_q=1$.
Since $\gamma_s$ (or $\gamma_s/\gamma_q$) 
controls the overall (relative)  yield of strange quarks,  one expects and 
finds in most  environments $\gamma_s\ne 1$ (or $\gamma_s/\gamma_q\ne 1$)  
and a value which  increases with system size, and  often with energy.

We recognize   considerable physics implication  of  understanding 
the value of $\gamma_q$, as  this relates directly to the measurement of  $T$,  
related to the phase  transformation condition $T_{\rm tr}$ of QGP to hadrons,
studied within lattice QCD. 
In the chemical equilibrium context,   $T\simeq T_{\rm tr}$. On the
other hand, the chemical non-equilibrium SHM implies rapid expansion and supercooled
transformation and hence $T< T_{\rm tr}$, with estimated difference 
at 10--15 MeV~\cite{Rafelski:2000by}.

For lower heavy ion  reaction energies as compared to LHC, one can 
also determine the baryochemical potential at hadronization. This then allows to compare
lattice QCD transformation condition,
$T_{\rm tr},\ \mu_{\rm B, tr}$~\cite{deForcrand:2010ys,Endrodi:2009sd}, with the data fit in the chemical equilibrium SHM~\cite{Andronic:2009gj,Cleymans:2010dw} and chemical 
non-equilibrium SHM~\cite{Letessier:2005qe}.
One finds that lattice results are much flatter compared to the 
equilibrium SHM, that is $T_{\rm tr}$ drops
off much slower with $\mu_{\rm B, tr}$. On the 
other hand, the non-equilibrium SHM parallels the lattice data 15 MeV below the 
transformation boundary.  This favors, on theoretical grounds, 
the chemical non-equilibrium approach.

{\bf LHC predictions assuming chemical equilibrium:}
Just one observable, the number of charged
particles $dh/dy\simeq 1800$, fixes within statistical model 
a large number of particle yields. This is allowing to  
test SHM  model. We explore chemical equilibrium and non-equilibrium 
conditions in turn.  We will  show that  many of the 
particle yields vary  little as function of hadronization condition 
for a prescribed $dh/dy$. We present an unusual number
of digits, a precision which has nothing to do with experiment but is needed to 
facilitate reproducibility of our results. We further state the propagation
error of the error $\Delta\, dh/dy=100$ which is mostly confined to the volume, 
but is in some cases also visible in  $T$ and also further below in the 
chemical non-equilibrium  parameters $\gamma_q,\ \gamma_s$,

Our procedure for the equilibrium SHM is as follows: in the chemical
equilibrium model, see   table \ref{eq_fit}, we set 
$\gamma_q=1,\ \gamma_s=1$ and  as noted also on other grounds $\lambda_q=1.0055$.
$\lambda_s$ follows from the  constraint of strangeness balance $s-\bar s =0$. 
All our results maintain a fixed $Q/B=0.4$. We then see that,
at fixed  hadronization temperature $T$ chosen in  table \ref{eq_fit} 
to be 159, 169, 179, 189 MeV, the yield of charged hadrons $dh/dy$ is, 
correlated to the  source  volume $dV/dy$.
Volume varies strongly with temperature, since particle yields scale, for $T\gg m$, 
as $VT^3$. Actually, since $T\gg m$  condition is not satisfied, the volume is
 changing more rapidly so that $T^kdV/dy={\rm Const.},\ k\simeq 7.2$.

Since charged particle number is fixed, we also expect that entropy of the bulk
is fixed and that is true; up to a small variation due to variation with $T$ 
in relative yield of heavy hadrons, the   entropy content is $dS/dy= 14800 \pm 400$.
As temperature increases, the proportion of heavy mass charged particles increases, 
and thus the pion yield and even the kaon yield slightly decrease with increasing $T$.
Baryon yields is most sensitive to $T$: $\Omega$ doubles in
yield in the temperature interval considered. Our choice of $\lambda_q$
fixes for each hadronization temperature the per rapidity net baryon 
yield also shown in table \ref{eq_fit}. --- We believe our choice is 
reasonable and has been made also so that we see that there is no  
need to distinguish particles from antiparticles. 

Yield of strangeness is slightly increasing, this increase is in heavy mass strange 
baryons, e.g., $\Lambda$, and this depletes slightly the yield of kaons.  Overall the specific
strangeness per entropy yield grows very slowly from $s/S=0.0245$ to 0.027. 
Several ratios, such as $\phi/{\rm K}^{0*}(892)\simeq 0.45 $, where several effects
compensate are nearly constant.

We also show the post-weak decay $\pi^0_{\rm WD}$ yield which is relatively 
large  and independent of hadronization $T$. The decay $\pi^0\to \gamma\gamma$ 
generates a strong electromagnetic energy  component.

\begin{table}\footnotesize
\caption{\label{eq_fit} 
Chemical equilibrium particle yields at $\sqrt{s_{NN}}= 2.76$ TeV. 
Top section: input  properties;
middle section: properties of  the fireball associated with central rapidity;
 bottom section: expected particle yields, and some select ratios.\newline 
* signals an input value, and ** a result directly following  from   input 
value (combined often with a constraint). 
All yields, but $\pi^0_{\rm WD}$, without   week decay feed 
to particle yields. Error in $dV/dy$ corresponds to error in $dh/dy$
 equal 100. }
\begin{ruledtabular}
\begin{tabular}{c| r | r | r |r }
\hline
$T^*$[MeV]               &$159$         & $169$       & $179$        & $189$          \\
$\gamma_{q}^*$           & $ 1$         &  $1$        &$  1$         & $  1$          \\
$\gamma_{s}^*$           & $1$          &   $1$       &$ 1$          & $  1$          \\
$\lambda_{q}^*$          & $ 1.0055$    &  $1.0055$   &$  1.0055$    & $  1.0055$     \\
$10^3(\lambda_{s}-1)^{**}$ & $2.06$       &  $1.45$     &$  0.89 $     & $ 0.39 $     \\
$(Q/B)^*$               &$0.4000$       &$0.4000$     &$0.4000$      &$0.4000$        \\
$(s-\bar s)^*$          &$0.0000$       &$0.0000$     &$0.0000$      &$0.0000$        \\
$(dh/dy)^*$               &$1800$         & $1800 $     & $1800$       & $1800$         \\
\hline
$dV/dy$ [fm$^3$]       &5285$\pm$147  &3452$\pm$96  &2286$\pm$64   &  1538$\pm$43   \\
$dS/dy$                &15155         & 14940       & 14690        &  14420         \\
$s/S$                  &$0.0245$      & 0.0255      & 0.0263       &  0.0270        \\
$P$ [MeV/fm$^3$]       &$64.1$        & 100         & 153          & 231            \\
$E/TS$                 &0.859         & 0.86        & 0.87         &  0.87          \\
$P/E$                  & 0.164        &0.158        & 0.153        &  0.150         \\
$E/V$ [GeV/fm$^3$]     &0.392         & 0.632       & 0.997        &   1.54         \\
\hline
$\pi^-$, $\pi^+$       &839           &830          & 821          &  813           \\
K$^-$                  &141.3         &140.8        & 139.0        & 136.8        \\
K$^+$                  &142.1         &141.6        & 140.1        & 137.8          \\
p                      &53.6          &63.1         & 72.0         & 79.8           \\
$\bar{\rm p}$          &51.9          &61.2         & 69.7         &  77.3          \\  
$\Lambda$              &30.0          &36.3         & 42.1         &  47.3          \\  
$\overline\Lambda$     &29.2          &35.4         & 41.1         &  46.2          \\  
$\Xi^-$                &4.45          &5.47         & 6.41         &  7.23           \\  
$\overline\Xi^+$       &4.36          &5.38         & 6.31         &  7.14           \\  
$\Omega^-$             & 0.772        &1.038        & 1.314        &  1.586         \\  
$(B-\overline B)^{**}$ &4.81          &  5.60       & 6.29         & 6.88           \\
$\rho$                 &92.4          &96.6         & 99.1         &  100.3          \\  
$\phi$                 &19.0          &20.5         & 21.4         &   21.9         \\  
K$^{0*}(892)$          &42.6          &45.3         &  46.9        &  47.6          \\  
K$^{0*}(892)$/K$^-$    &0.301         &0.322        &  0.337       &  0.348         \\ 
$\phi$/K$^{0*}(892)$   &0.446         & 0.452       & 0.456        &   0.460        \\ 
$\pi^0$                &942           &933          & 925          &   916          \\  
$\eta$                 &110           &111          & 111          & 110          \\  
$\eta'$                &9.67          &10.4         & 10.8         &  11.1           \\  
$\pi^0_{\rm WD}$       &1251          &1251         & 1249         &  1245          \\  
\hline
\end{tabular}
\end{ruledtabular}
\end{table}

\begin{table}\footnotesize
\caption{\label{neq_fit} 
Chemical \underline{non}-equilibrium particle yields, each column for different hadronization pressure. See caption of table \protect\ref{eq_fit} for further details.} 
\begin{ruledtabular}
\begin{tabular}{c| r | r | r | r  }
\hline
$P^*$ [MeV/fm$^3$]      &$60.3 $     &$70.0 $      &$82.2 $     &$90.1 $    \\
$(s/S)^*$               &$0.0367 $   &$0.0370  $   &$0.0370$    &$0.0373 $    \\
$\lambda_{q}^*$         & $ 1.0055$    &  $1.0055$   &$  1.0055$    & $  1.0055$     \\
$10^3(\lambda_{s}-1)^{**}$   & $2.69$       &  $2.45$     &$  2.19 $     & $2.04   $     \\
$(Q/B)^*$               &$0.400$     &$0.400$       &$0.4000$      &$0.4000$     \\
$(s-\bar s)^*$          &$0.0000$    &$0.0000$      &$0.0000$      &$0.0000$     \\
$(dh/dy)^*$             &$1800$      &$1800$        &$1800$        &$1800$      \\
\hline
$T$ [MeV]               &$131.2$    &$134.3\pm0.1 $   &$137.7\pm0.1$  & $139.6\pm0.1$      \\
$\gamma_{q}$         &$1.599 $   &$1.600 $     &$1.601 $    & $1.599 $  \\[-0.1cm]
                     &$ \pm0.001$  &$ \pm0.008$  &$ \pm0.009$ & $ \pm0.011$  \\ 
$\gamma_{s}$         &$2.913$    &$2.842$   &$2.745$    & $2.721$ \\[-0.1cm]
                     &$\pm0.008$ &$\pm0.030$&$\pm 0.030$& $\pm 0.016$ \\[0.1cm]
$dV/dy$ [fm$^3$]     &$5469$$\pm542$    &$4731$$\pm136$    &$4043$$\pm119$    &$3705$$\pm168$  \\
$dS/dy$                 &13924        &13879         &13794         & 13797          \\
$E/TS$                  &1.060        &1.060         &1.059         & 1.059          \\
$P/E$                   &0.170        &0.168         & 0.165        &0.164           \\
$E/V$ [GeV/fm$^3$]      &0.354        &0.417         &0.497         &0.550           \\
\hline
$\pi^-$, $\pi^+$        &858          &854           &850           &848             \\
K$^-$                   &192.0        &190.2         &186.5         &186.2           \\
K$^+$                   &192.9        &191.2         &187.5         &187.3           \\
p                       &32.9         &36.3          &40.3          & 42.5           \\  
$\bar{\rm p}$           &31.8         &35.2          &39.0          &41.2            \\  
$\Lambda$               &28.9         &31.8          &34.8          &36.9            \\  
$\overline\Lambda$      &28.1         &31.0          &33.9          & 35.9           \\  
$\Xi^-$                 &6.92         &7.56          &8.12          &8.60            \\  
$\overline\Xi^+$        &6.77         &7.40          &7.96          &8.43            \\  
$\Omega^-$              &1.56         &1.73          &1.89          &2.03            \\  
$(B-\overline B)^{**}$  &3.640        &3.973         &4.328         & 4.539          \\
$\rho$                  &56.1         &58.8          &61.7          &63.2            \\  
$\phi$                  &30.0         & 30.7         &30.8          &31.4            \\  
K$^{0*}(892)$           &39.9         &41.4          &42.6          &43.6            \\  
K$^{0*}(892)$/K$^-$     &0.208        &0.218         &0.228         &0.234           \\ 
$\phi$/$K^{0*}(892)$    &0.751        &0.741         &0.722         & 0.721          \\ 
$\pi^0$                 &988          &983           &979           & 977            \\  
$\eta$                  &134          &132           &128           &128             \\  
$\eta'$                 &10.4         &10.7          &10.8          &11.0            \\  
$\pi^0_{\rm WD}$        &1398         &1396          & 1389         & 1391          \\  
\hline
\end{tabular}
\end{ruledtabular}
\end{table}

{\bf LHC prediction within non-equilibrium SHM:}
Within   the  non-equilibrium hadronization approach, we need to 
further  anchor the two quark pair abundance parameters  
$\gamma_q$ and $\gamma_s$. In absence of experimental data,
we introduce additional hadronization conditions, the relative 
strangeness yield $s/S$ and hadronization pressure $P$. 
 We  vary $P$, see table \ref{neq_fit},  just as we varied  
 $T$  the hadronization condition
in the chemical equilibrium model. In the chemical non-equilibrium 
SHM, two conditions suffice to 
narrow considerably the values of three SHM parameters 
($\gamma_q$, $\gamma_s$ and $T$), but only if we insist
that a third condition $E/TS>1$ is qualitatively satisfied. 

Strangeness yield is a natural hadronization condition of QGP.
We consider the ratio of strangeness per entropy 
$s/S$ in which $T^3$ coefficients
and other systematic dependencies cancel.  Since the entropy
contents is directly related to the particle multiplicity $dh/dy$,
in our case $s/S$ implies $s$-yield and directly relates to strangeness
pair yield. In QGP,
$s\bar s$-pairs are  produced predominantly in thermal gluon processes and 
their yield can be obtained within the QCD perturbative approach. 
In a study which was refined to agree with the strangeness
yield observed at RHIC, we  predicted the value 
$s/S\simeq 0.037$ for LHC~\cite{Letessier:2006wn}. We use
here this result, noting that   higher $s/S$ values
are possible, depending on LHC formed QGP dynamics.
 The QGP expected dynamic strangeness yield is 
considerably higher than the chemical 
equilibrium yield, table \ref{eq_fit}, $0.0270\le s_{\rm eq}/S \le 0.0245$. The 
greater strangeness content in QGP is, in fact, the reason behind the interest
in strangeness as signature of QGP. 
 
The second condition arises from the observation that  once the 
statistical parameters were  fitted across
diverse reaction conditions at  RHIC, the one constant 
outcome was  that  the hadronization 
pressure $P= 82$ MeV/fm$^3$~\cite{Rafelski:2009jr}.
Choice of   pressure as a natural QGP hadronization constraint is
further rooted in the observation that the vacuum confinement phenomenon
can be described within the qualitative 
MIT-bag model of hadrons introducing  vacuum pressure is $B_{\rm MIT}=58$ MeV/fm$^3$,
while in a bag-motivated fit to hadron spectra which allows additional flexibility 
in parameters one finds $B_{\rm fit}=112$  MeV/fm$^3$~\cite{Aerts:1984vv}.
Clearly, a range of values is possible theoretically, with the hadronization 
condition $P=82$ MeV/fm$^3$ right in the middle of this domain. 
 We will use this `critical pressure' hadronization condition
as our  constraint, but also vary it such that $60  \le P\le 90$ MeV/fm$^3$
so that we can be sure that our prediction is not  critically dependent on 
the empirical value. 
The pressure seen in equilibrium model case, table \ref{eq_fit}, has a 
range $64\le P\le 230$ MeV/fm$^3$. 

The third constraint   is not imposed in its precise value, 
but we require that   hadronization occurs 
under the constraint that  $E/TS>1$. In comparison, for 
the equilibrium case, table \ref{eq_fit}, we have $0.86<E/TS<0.87$ a relatively small
variation. The importance of this quantity $E/TS>1$ as a diagnostic tool for explosive 
QGP outflow and hadronization was discussed in~\cite{Rafelski:2000by}. In fact, we find
that a reasonable and stable hadronization arises in chemical non-equilibrium
 within a narrow interval $1.059<E/TS<1.060$.

In table \ref{neq_fit}, the outcome   of this procedure is presented. We state
the actual values of parameters for which solution of all constraints was numerically
obtained, thus in first column pressure is not 60 but 60.3 MeV/fm$^3$. 
With rising hadronization pressure, the hadronization temperature rises, but it remains
well below the phase transformation temperature. As we have discussed, the low value of $T$
in the chemical non-equilibrium SHM is consistent with the  
dynamics of the expansion, the flow of matter reduces the phase balance $T$. This in
turn is then requiring that the light quark abundance parameter $\gamma_q\simeq 1.6$. 
This is the key distinction of the chemical non-equilibrium.  It further signals enhancement
of production of baryons over mesons by just this factor. 
Note that $\gamma_s/\gamma_q\simeq 1.72$. This large ratio means that   
the yield of $\Lambda$ and $p$ do  not differ much. This indicates  strong enhancement 
of strangeness, a first day observable of QGP formation, 
along with $\phi$ enhancement~\cite{Rafelski:1982ii}.

{\bf Comparison of SHM results:}
The large bulk hadronization 
volume  $dV/dy\simeq 4500$\,fm$^3$  is   suggesting that there will 
 be noticeable changes in the HBT observables allowing to produce
such a great hadronization volume. The bulk energy content is found  
in both approaches  to be $dE/dy=E/V\times dV/dy= 2.00\pm 0.05$ TeV
per unit of rapidity at $y=0$. This is the thermal energy
of QGP prior to hadronization measured in the local fluid element rest frame. 

The entropy content in the bulk for non-equilibrium, $dS/dy=13860\pm64$, 
is 5\% smaller compared 
to equilibrium case. This is   due to the fact that non-equilibrium 
particle yields do not maximize entropy. We note that the yields of pions,
kaons, and even single strange hyperons are remarkably independent of hadronization 
pressure, or, in the equilibrium case, temperature, 
even though the volume parameter changes greatly. This effect is 
counteracted by a balancing change in hadronization temperature since the yield of 
charged hadrons is fixed.
The hadronization energy density is very close to $E/V\simeq 0.5$ GeV/fm$^3$,
and it tracks the pressure since the ratio $P/E$ is found to be rather constant. 

\begin{figure}[h]
\centerline{\hspace*{-0.3cm}\includegraphics[width=2.9in]{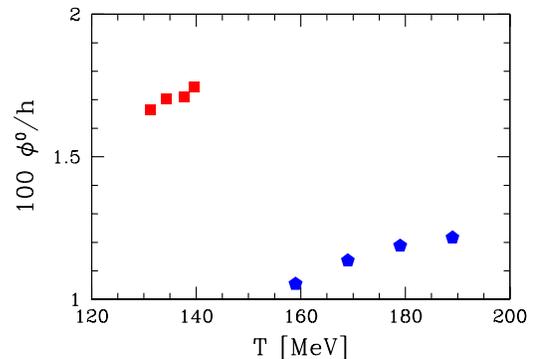}}
\caption{\label{phih} (color on-line)  Specific yield $\phi/h$ for chemical
equilibrium model to right bottom (blue) and chemical non-equilibrium model 
left up (red).
}
\end{figure}

The yield of multistrange hadrons  is much enhanced in chemical 
non-equilibrium model, compared to the equilibrium model, 
on account of 50\% increased yield of
strangeness, which is potentiated for multistrange particles as 
was predicted to be the signature of QGP~\cite{Rafelski:1982ii}. 
The yield of $\Lambda$, for the most favored 
hadronization condition in both equilibrium and non-equilibrium, 
can in fact be lower in the non-equilibrium case
than in the equilibrium, yet at LHC the yield of K is always 
40\% greater. Like in the equilibrium results, we observe
that several yields are largely independent of the hadronization
condition, meaning that ratios such as $\phi/h$ could be 
a distinctive signature of hadronization, differentiating 
the two primary models. This is illustrated in figure \ref{phih} 
where the ratio is shown as function of resultant hadronization $T$.

\begin{figure}[t]
\centerline{\hspace*{-0.3cm}\includegraphics[width=2.8in]{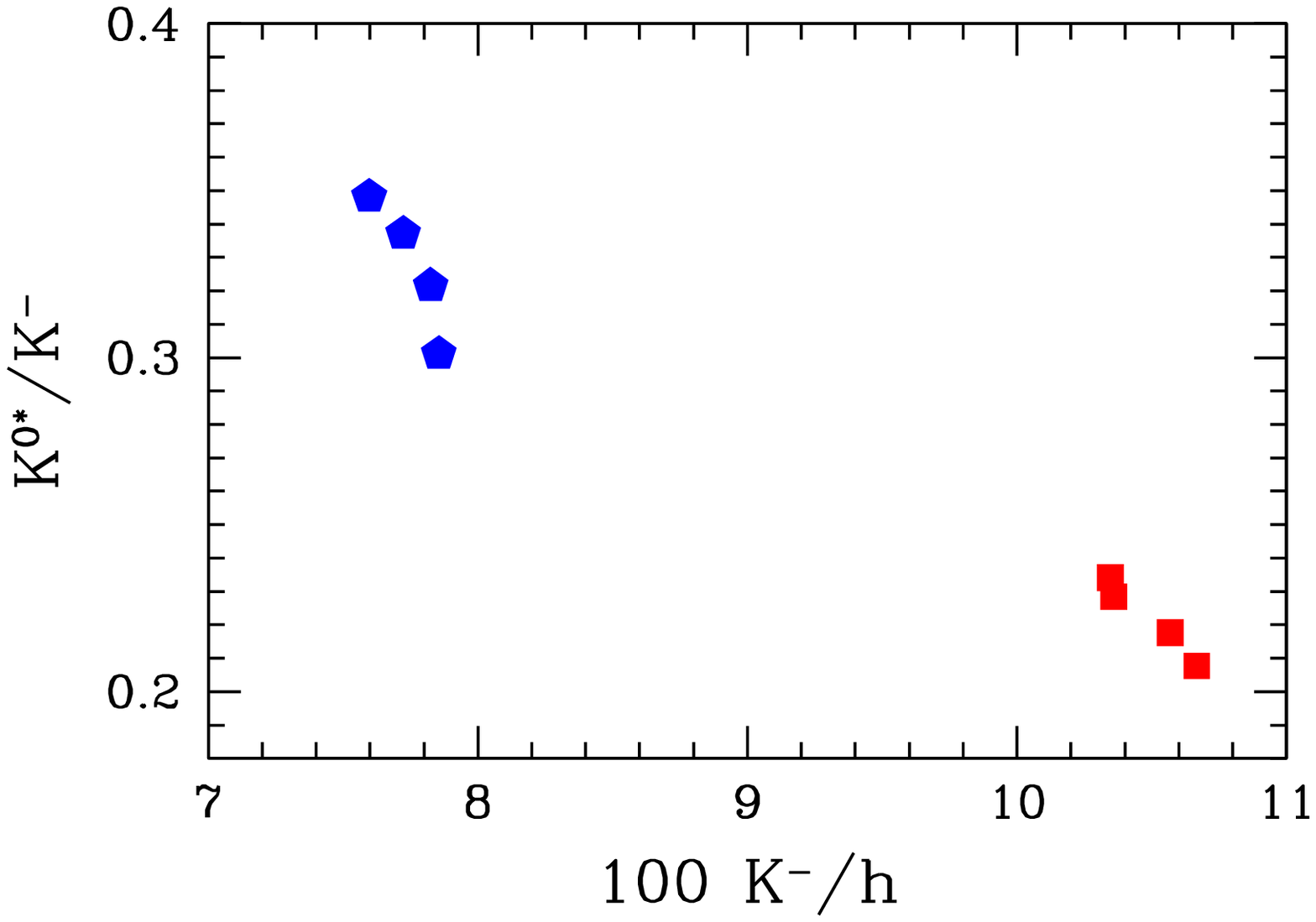}}
\centerline{\hspace*{-0.3cm}\includegraphics[width=2.8in]{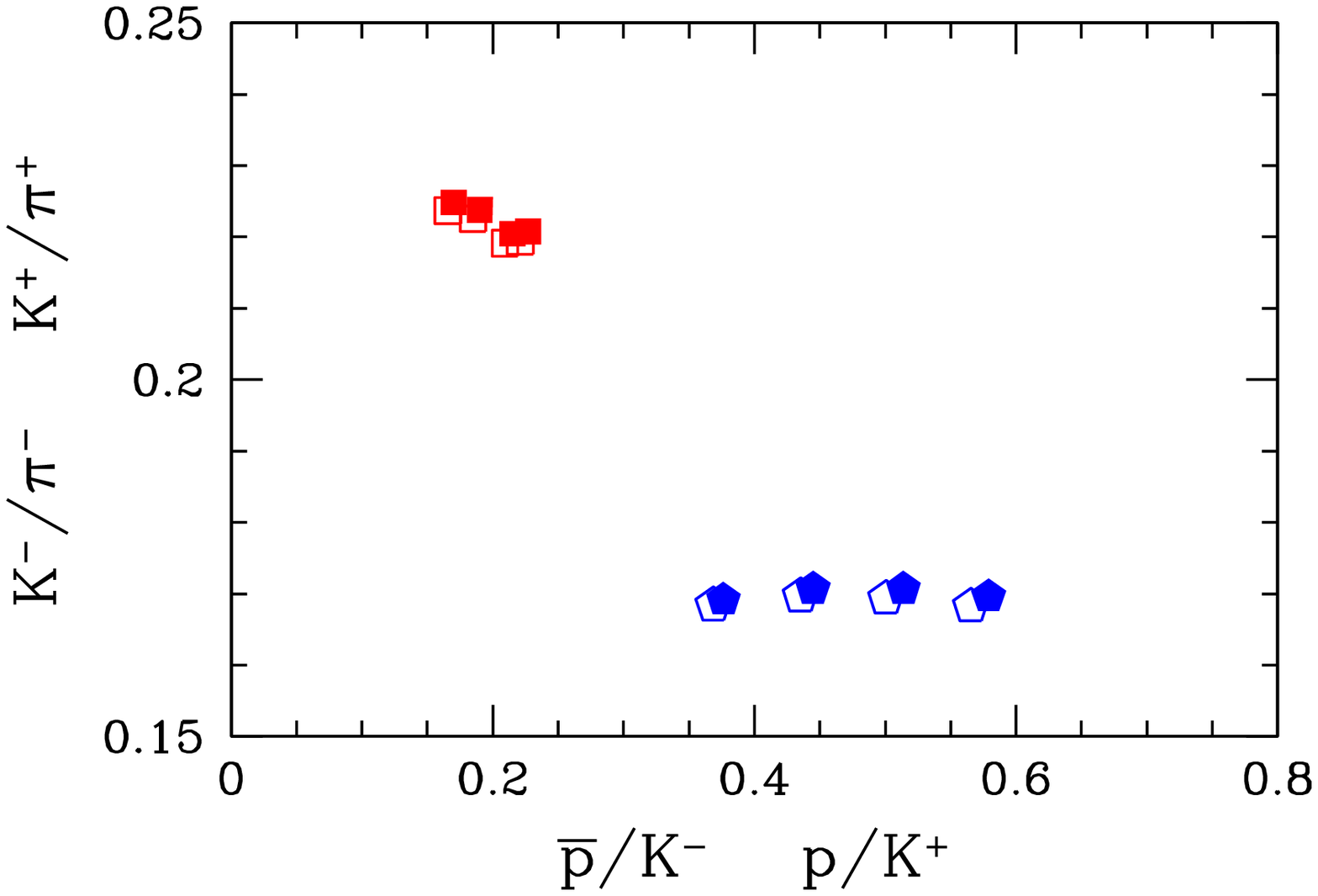}}
\caption{\label{KstarKpi} (color on-line) 
Top: ratio of resonance K$^*$ to kaon yield K, as function of specific 
kaon yield K$^-/h$, the equilibrium model (blue) is left up
and chemical non-equilibrium (red) is bottom right. 
Bottom frame: the K$/\pi$ ratio   
as function of $p/K$ ratio. Bottom right (blue) is equilibrium 
model and upper left is non-equilibrium model.
}
\end{figure}

One can, however, argue that  $\phi/h$,  seen in  Fig.\,\ref{phih}, could be
brought about by strangeness enhancement, not requiring that $\gamma_q>1$,
for a more complete discussion see~\cite{Petran:2009dc}. To narrow the choices,
we propose to study two more ratios,  shown in Fig.\,\ref{KstarKpi}. The top
section shows ${\rm K}^*/{\rm K}^-$,  as function of the yield ${\rm K}^-/h$.
The lower right (red) non-equilibrium result shows strangeness enhancement at low
hadronization $T$ since  ${\rm K}^*/{\rm K}^-$ mainly depends on $T$. 
We have shown both particle and antiparticle ratios derived from our fixed 
input for net baryon yield to illustrate that differentiation of these results 
will not be easily possible. 

The bottom section, in Fig.  \ref{KstarKpi}, shows the strangeness enhancement 
in format of ${\rm K}/\pi$  as function of the easiest to measure baryon to 
meson ratio which is  a measure of absolute magnitude of $\gamma$,
here specifically,   $\gamma_q^3/\gamma_s\gamma_q$. We see  the equilibrium model 
to lower right (blue) while the non-equilibrium model is upper left.

{\bf Summary and conclusions:}
We have obtained particle yields within the statistical 
hadronization model for the 
LHC-ion run at  $\sqrt{s_{\rm NN}}  = 2.76$ TeV. We have discussed both
bulk properties of QGP at breakup and the resulting particle yields. 
These can vary significantly depending on the   hadronization mechanism.
Distinctive features associated with QGP based strangeness 
enhancement and final state  chemical non-equilibrium
were described  and strategies leading to 
better understanding of chemical (non-)equilibrium
were proposed. 

We have shown how   enhanced yield of strangeness
with the phase space occupancy $\gamma_s\simeq 2.75$ when $\gamma_q\simeq 1.6$
modifies the yield of strange hadrons and detailed predictions for the observables
such as $\phi/h$, ${\rm K}^*/{\rm K}$, ${\rm K}/\pi$, $p/{\rm K}$, $\Lambda/p$ were offered.
Enhanced yields of (multi)strange particles are tabulated. 
We note that absolute yield of $\phi$ is enhanced by a factor 1.5 
in the non-equilibrium compared to equilibrium hadronization. There is no significant dependence
of the $\phi$ yield on hadronization condition making it an ideal first day  differentiating
chemical equilibrium from non-equilibrium.  

The large bulk hadronization volume  
$dV/dy\simeq 4500$\,fm$^3$  related to HBT observables, the local rest frame
thermal energy content  $dE/dy|_0=2$ TeV constrains hydrodynamic models.  
A large yield of $\pi^0$, $\eta$ and thus of associated 
decay photons is noted, enhanced somewhat  in the.chemical non-equilibrium case.

{\bf Acknowledgments} LPTHE: Laboratoire de Physique Th{\' e}orique 
et Hautes Energies,  at University Paris
6 and 7 is supported by CNRS as Unit{\' e} Mixte de Recherche, UMR7589.
 This work was supported by a grant from the 
U.S. Department of Energy, DE-FG02-04ER41318

%

\end{document}